\documentclass[sigconf]{acmart}

\AtBeginDocument{%
  \providecommand\BibTeX{{%
    \normalfont B\kern-0.5em{\scshape i\kern-0.25em b}\kern-0.8em\TeX}}}

\copyrightyear{2022}
\acmYear{2022}
\setcopyright{rightsretained}
\acmConference[WSDM '22]{Proceedings of the Fifteenth ACM International Conference on Web Search and Data Mining}{February 21--25, 2022}{Tempe, AZ, USA}
\acmBooktitle{Proceedings of the Fifteenth ACM International Conference on Web Search and Data Mining (WSDM '22), February 21--25, 2022, Tempe, AZ, USA}
\acmDOI{10.1145/3488560.3498454}
\acmISBN{978-1-4503-9132-0/22/02}
\usepackage{tikz}
\usetikzlibrary{arrows.meta}
\DeclareMathOperator{\diag}{diag}

\DeclareMathOperator{\corr}{corr}
\DeclareMathOperator{\prob}{Prob}
\DeclareMathOperator{\sign}{Sign}
\DeclareMathOperator{\walk}{Walk}
\DeclareMathOperator*{\mean}{mean}

\DeclareMathOperator{\vol}{vol}
\usepackage{mathtools}
\DeclarePairedDelimiter\norm{\lVert}{\rVert}%
\DeclareMathOperator{\rank}{rank}
\DeclareMathOperator*{\argmin}{arg\,min}

\newtheorem{property}{Property}
\usepackage[normalem]{ulem}
\usepackage{multirow}
\usepackage{float}
\usepackage{subfig}
\usepackage{enumitem}
\usepackage{multicol}
\begin{document}

\title{POLE: Polarized Embedding for Signed Networks}
\author{Zexi Huang}
\affiliation{%
 \institution{University of California}
 \city{Santa Barbara}
 \state{CA}
 \country{USA}
}
\email{zexi_huang@cs.ucsb.edu}

\author{Arlei Silva}
\affiliation{%
 \institution{Rice University}
 \city{Houston}
 \state{TX}
 \country{USA}}
\email{arlei@rice.edu}

\author{Ambuj Singh}
\affiliation{%
 \institution{University of California}
 \city{Santa Barbara}
 \state{CA}
 \country{USA}}
\email{ambuj@cs.ucsb.edu}

\renewcommand{\shortauthors}{Huang, et al.}

\begin{abstract}
From the 2016 U.S. presidential election to the 2021 Capitol riots to the spread of misinformation related to COVID-19, many have blamed social media for today's deeply divided society. Recent advances in machine learning for signed networks hold the promise to guide small interventions with the goal of reducing polarization in social media. However, existing models are especially ineffective in predicting conflicts (or negative links) among users. This is due to a strong correlation between link signs and the network structure, where negative links between polarized communities are too sparse to be predicted even by state-of-the-art approaches. To address this problem, we first design a partition-agnostic polarization measure for signed graphs based on the signed random-walk and show that many real-world graphs are highly polarized. Then, we propose POLE (\underline{POL}arized \underline{E}mbedding for signed networks), a signed embedding method for polarized graphs that captures both topological and signed similarities jointly via signed autocovariance. Through extensive experiments, we show that POLE significantly outperforms state-of-the-art methods in signed link prediction, particularly for negative links with gains of up to one order of magnitude. 

\end{abstract}



\maketitle
\section{Introduction}

Social media has made our world more polarized \cite{tucker2018social,gillani2018me,garimella2017long}. The events surrounding the 2016 U.S. election \cite{guess2018selective} and, more recently, the tragic U.S. Capitol riot \cite{prabhu2021capitol} and spread of COVID-19 misinformation \cite{naeem2021exploration,m2021political}, have illustrated the dangers of a deeply ideologically divided society. But if technology has led to the rise of polarization, can it also help us to solve it? More specifically, can recent advances in representation learning \cite{perozzi2014deepwalk,javari2020rose} help us to address online polarization? One could argue these methods should be as effective for predicting conflicts as they are for recommending connections and content in social media platforms \cite{goyal2018graph}. However, we will show that polarization leads to new challenges for representation learning.

\begin{figure}
    \centering
    \subfloat[LFR-polarized]{ \includegraphics[width=0.20\textwidth]{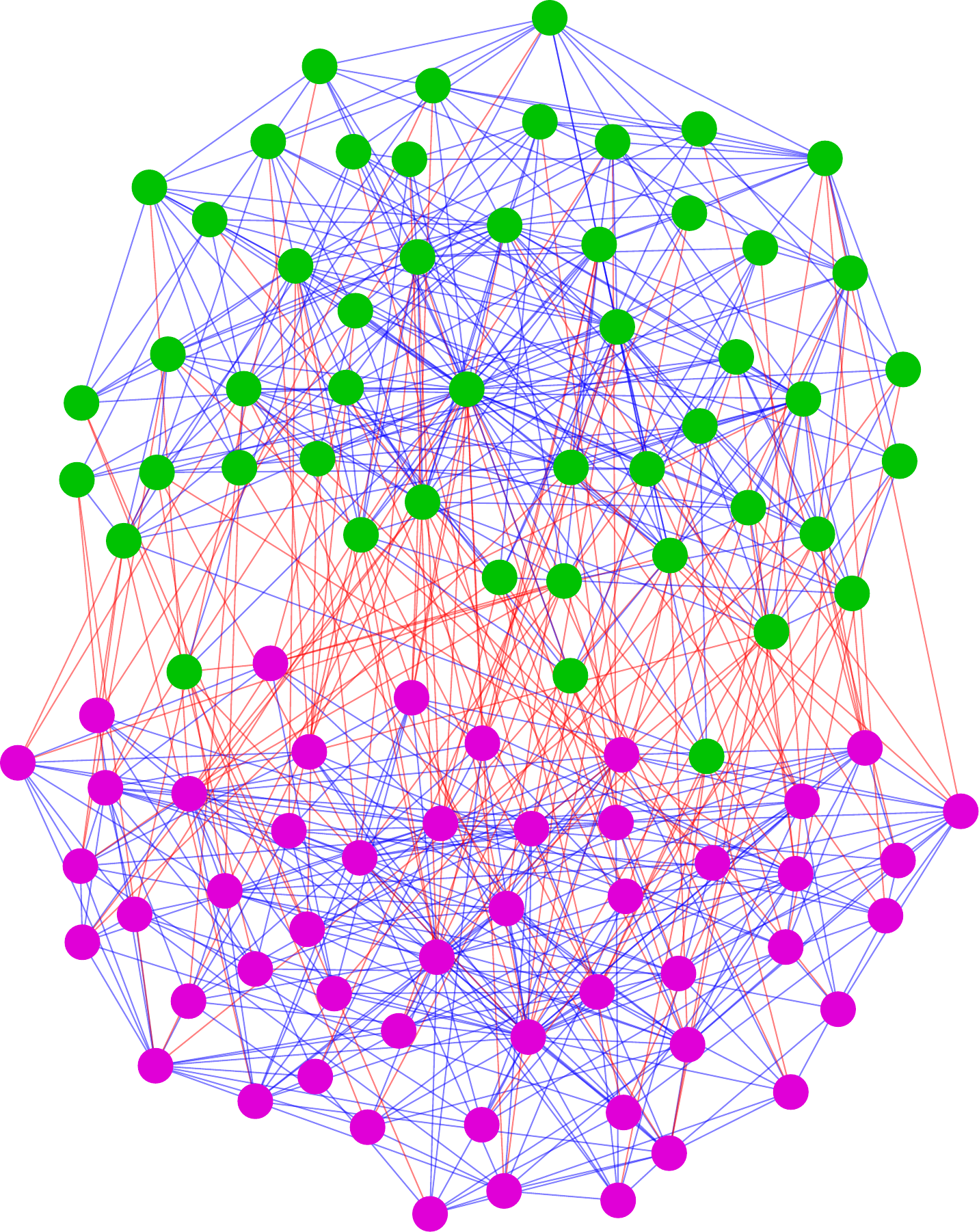}}
    \qquad
    \subfloat[LFR-unpolarized]{ \includegraphics[width=0.20\textwidth]{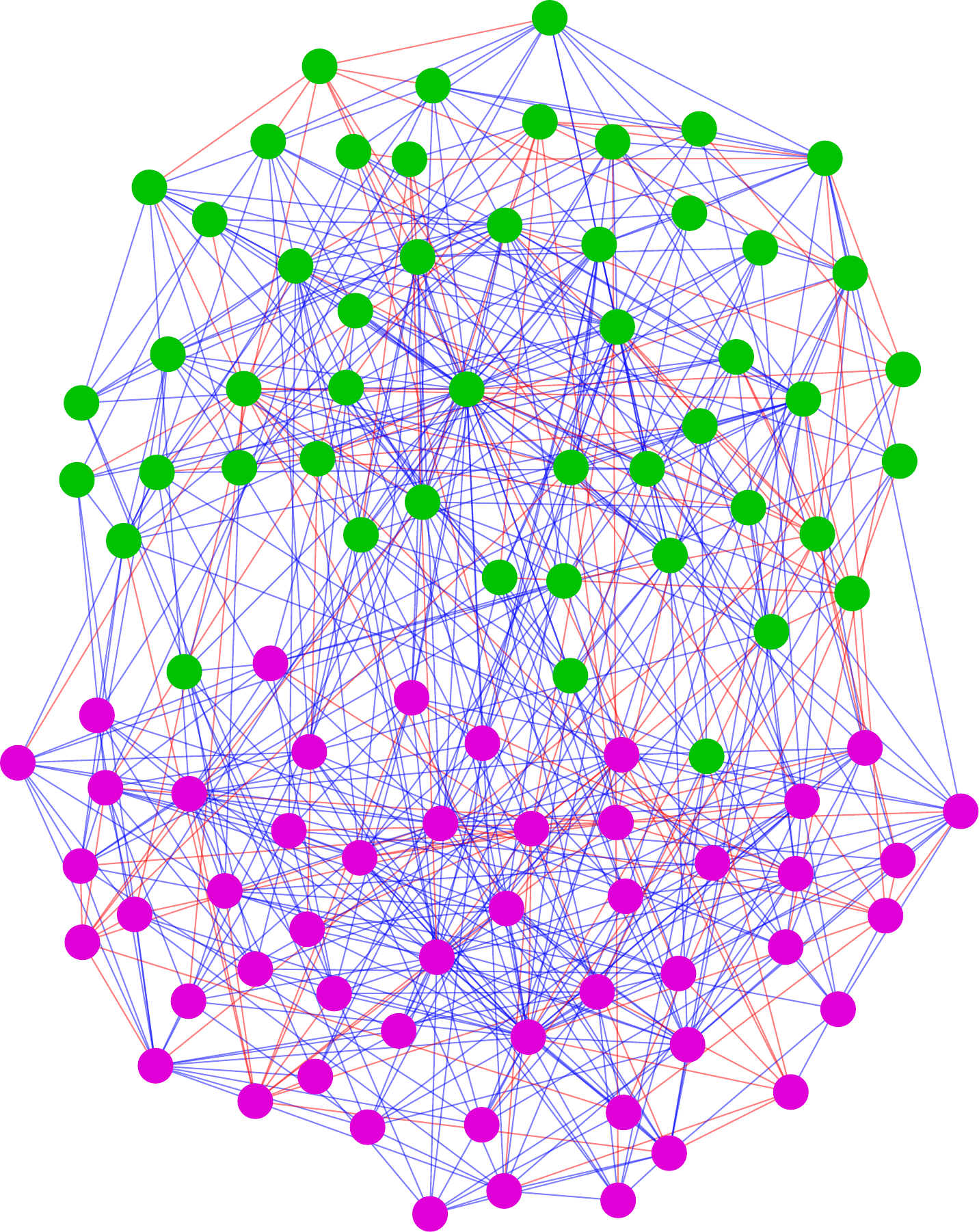}}
    \caption{Two synthetic graphs with the same underlying topology but different link signs. Node colors depict communities and link colors depict signs. (a) is more polarized than (b) as its link signs are related to the community structure---negative links connecting two polarized communities. }
    \label{fig::lfr}
\end{figure}

Signed graphs are a powerful tool for analyzing social polarization, capturing both positive (friendly) and negative (hostile) connections between entities. They have been used to model relationships between politicians in the U.S. Congress \cite{thomas2006get} and interactions between Twitter users on political matters \cite{lai2018stance}, both of which are known to harbor polarization \cite{theriault2006party, conover2011political}. In signed graphs, polarization is often related to the emergence of conflicting communities \cite{bonchi2019discovering}, where nodes within each community form dense positive connections and nodes across different communities are sparsely connected via negative links. \autoref{fig::lfr} shows two signed graphs with the same underlying topology (based on the LFR benchmark \cite{lancichinetti2008benchmark}) but different link signs. The left graph is more polarized as all of its negative links connect the two communities, while link signs in the right graph are unrelated to the community structure. 
 
To fight polarization, we first need to measure it. Existing measures \cite{bonchi2019discovering, tzeng2020discovering} are defined as objective functions for community detection, requiring community memberships as input. This dependence makes those measures less useful for real-world graphs, where ground-truth memberships are often unknown. To solve this problem, we propose a novel polarization measure based on the correlation between unsigned and signed random-walk dynamics. By varying the random-walk length, our measure naturally captures polarized community structure at different scales and does not rely on specific partitions. We will demonstrate its effectiveness in characterizing both node and graph-level polarization.

But how can we fight polarization in a network? For unsigned graphs, one approach is to bridge communities to reduce the effect of echo chambers \cite{garimella2017reducing}. Signed graphs open new possibilities in this endeavor. In particular, if one can identify potential hostile links that are likely to be formed in the future, they can take preventive measures to reduce further polarization. The success of this approach relies on the accuracy of signed link prediction \cite{chiang2011exploiting, hsieh2012low, wang2017online}, where both existence and signs of future links are inferred. However, while unsigned and signed graph embedding methods have shown success at the related link-level tasks of (unsigned) link prediction \cite{grover2016node2vec, random-walk-embedding} and sign prediction \cite{wang2017signed, kim2018side}, they cannot be easily combined for signed link prediction in polarized graphs. That is because negative links mostly connect antagonistic communities and are much sparser than positive links, making it nearly impossible to predict their existence with unsigned embedding methods. And without link existence information, even a perfect sign prediction oracle is unable to predict the negative links. 

To address the challenges mentioned above, we propose POLE, a novel signed embedding method for polarized graphs. The key feature that distinguishes our method from existing ones is that it captures signed and topological similarities jointly. Specifically, it guarantees that positively related pairs are more similar than unrelated topologically distant pairs, which are in turn more similar than negatively related pairs. This is accomplished by leveraging the signed random-walk that incorporates social balance theory \cite{heider1946attitudes} and extending autocovariance similarity \cite{delvenne2010stability, schaub2019multiscale} to signed graphs. In this way, negative links can be predicted as the most dissimilar node pairs in the graph, at the other end of the similarity spectrum. 

To summarize, our main contributions are:
\begin{itemize}
    \item We design a novel partition-agnostic polarization measure for signed graphs based on the signed random-walk. 
    \item We analyze how existing signed embedding methods fail in signed link prediction for polarized graphs. 
    \item We propose POLE, a novel signed embedding approach for polarized graphs that captures both topological and signed similarities via signed autocovariance. 
    \item We conduct an extensive experimental evaluation of our method on six real-world signed graphs. Results show that POLE significantly outperforms state-of-the-art methods in signed link prediction, especially for negative links.  
\end{itemize}
\section{Random-walk on Signed Graphs}
We introduce our formulation for random-walks on signed graphs, the basis for our polarization measure and embedding algorithm. 
\label{sec::random-walk}
\subsection{Notations}
A signed undirected weighted graph is a tuple $G=(\mathcal{V}, \mathcal{E})$, where $\mathcal{V}=\{1, \ldots n\}$ denotes the set of $n$ nodes and $\mathcal{E}=\mathcal{E}_+\cup \mathcal{E}_-$ denotes the set of $m$ links, with positive and negative signs. The graph is represented by a signed weighted adjacency matrix $A\in \mathbb{R}^{n\times n}$, with $A_{uv}>0$ (or $<0$) if a positive (or negative) link of weight $|A_{uv}|$ connects nodes $u$ and $v$, and $A_{uv}=0$ 
otherwise. We also use the absolute adjacency matrix $|A|$ to construct the degree matrix/vector $D$/$d$ with $D_{uu}=d_u=\sum_{v}|A|_{uv}$ as the degree of node $u$. 

\subsection{Signed Random-walk}
\label{subsec::signed_random_walk}
Random-walks are widely used for unsigned graph embedding due to their ability to capture topological similarity at multiple structural scales. While there have been several attempts to extend random-walks to signed graphs \cite{jung2016personalized, yin2021signedpagerank}, we will define our own version to capture topological and signed similarity jointly and explicitly guarantee \emph{polarized similarity consistency}---we will discuss this important property in more detail in the next subsection.

We start by recalling random-walks for unsigned graphs. A walk $l$ is a sequence of nodes $\langle w_0, w_1, \ldots w_t \rangle$ where $(w_\tau,w_{\tau+1}) \in \mathcal{E}$. The transition probability of the walk is the product of stepwise ones:
\begin{equation}
\begin{aligned}
        \prob(l) &= \prod_{(w_\tau,w_{\tau+1})\in l}\prob(w_{\tau+1}|w_\tau)\\
        &=\prod_{(w_\tau,w_{\tau+1})\in l}|A|_{w_{\tau}w_{\tau+1}}/d_{w_{\tau}}
\end{aligned}
\end{equation}
Then, the $t$-step random-walk transition probability from node $u$ to $v$ can be expressed as the sum of the transition probabilities of all length-$t$ walks between $u$ and $v$, denoted as $\walk(u,v,t)$:
\begin{equation}
    |M|_{uv}(t)= \sum_{l \in \walk(u, v; t)} \prob(l)
\end{equation}
which serves as a measure of topological similarity between $u$ and $v$. The Markov time $t$ controls the scale of the walk. 

\begin{figure}[htbp]
    \centering
    \includegraphics[width=1\columnwidth]{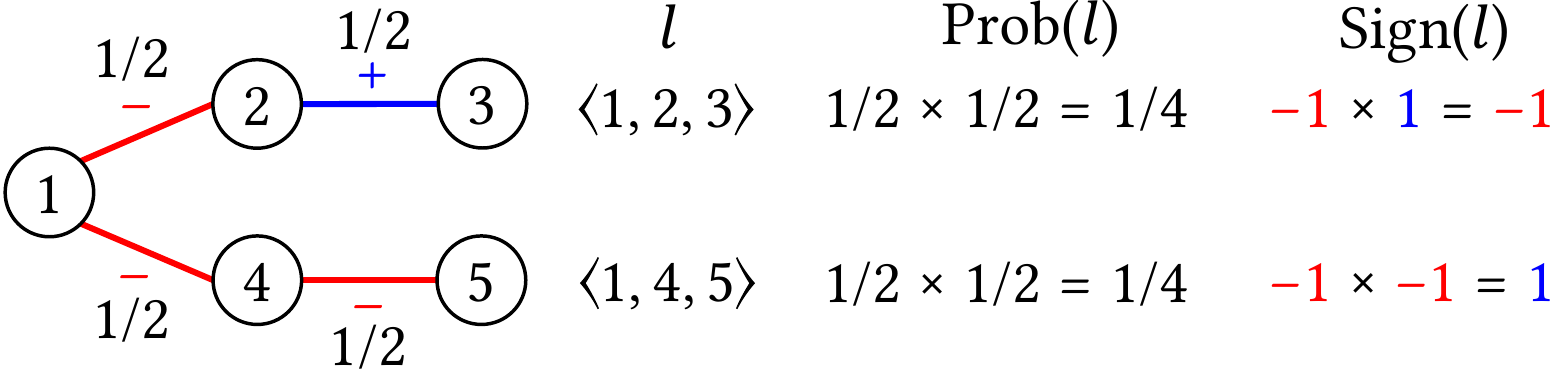}
    \caption{Examples of signed walks with corresponding transition probabilities ({\normalfont $\text{Prob}$}) and inferred signs ({\normalfont $\text{Sign}$}).}
    \label{fig::signed-path}
\end{figure}

For signed graphs, we keep transition probabilities of walks for topological similarity and add an inferred sign for each walk to capture signed similarity, leading to
\begin{equation}
    M_{uv}(t) = \sum_{l \in \walk(u, v; t)} \sign(l)\prob(l)
    \label{eqn::signed-transition}
\end{equation}
where $\sign(l)=\sign(\prod_{(w,w') \in l}A_{ww'})$ determines the sign of the walk $l$ between $u$ and $v$. We leverage the well-established social balance theory \cite{heider1946attitudes} to infer the sign of the walk, which states the famous rule that ``an enemy of my enemy is my friend'' among other rules. \autoref{fig::signed-path} shows examples of signed transitions.

Note that after adding signs to walks, $M_{uv}(t) \in [-1, 1]$ is no longer a probability but rather captures a notion of signed similarity between $u$ and $v$. In matrix form, we will still call $M(t) \in \mathbb{R}^{n\times n}$ the signed random-walk transition matrix even though it is not stochastic. A nice property of $M(t)$ is that it can be conveniently expressed in terms of the signed adjacency and degree matrices:
\begin{equation}
    M(t)\!=\!\begin{cases}
	(D^{-1}A)^t & \text{for discrete random-walks}\\
	\exp (-(I-D^{-1}A)t) & \text{for continuous random-walks}
	\end{cases}
\end{equation}
where $I \in \mathbb{R}^{n\times n}$ is the identity matrix. We will use the continuous version of the transition matrix in the rest of the paper due to its finer granularity of the Markov time ($t\in [0, +\infty)$ instead of $t\in \mathbb{N}_0$).

\subsection{Similarity Consistency}
\label{subsec::similarity_consistency}
The key advantage of our signed random-walk is that it guarantees \emph{polarized similarity consistency}, a property critical to signed link prediction in polarized graphs. The \emph{consistency} of similarity states which type of node pairs should be more similar than other types. For example, unsigned embedding methods satisfy:
\begin{property}[Topological similarity consistency]
Topologically close nodes are more similar than topologically distant ones.

\end{property}
\noindent And signed embedding methods guarantee: 
\begin{property}[Signed similarity consistency]
Positively related nodes are more similar than negatively related ones.
\end{property}
\noindent These properties enable corresponding link-level downstream tasks, namely (unsigned) link prediction and sign prediction. 

The transitions of our signed random-walk as a similarity metric have a stronger consistency guarantee: 
\begin{property}[Polarized similarity consistency]
Positively related node pairs are more similar than unrelated topologically distant pairs, which are in turn more similar than negatively related pairs. 
\label{prop::three}
\end{property}
\noindent Property \ref{prop::three} is a direct implication of the transition probability defined in \autoref{eqn::signed-transition}. Large positive/negative values of $M_{uv}(t)$ indicate that $u$ and $v$ are connected via mostly positive/negative paths while small values mean that they are unrelated and topologically distant. Property \ref{prop::three} also holds for a similarity defined by dot products between columns of $M$. More specifically, $\langle M_{:u}, M_{:v}\rangle$ is large and positive (or negative) if and only if $u$ and $v$ have transitions of same (or different) sign(s) to a common set of nodes. And the similarity is small if their transitions reach a distinct set of nodes.

\section{A Measure of Polarization}

In this section, we propose and evaluate a measure of polarization based on the signed random-walks we have just introduced.

\subsection{Random-walk Based Polarization}
A signed network is polarized if it comprises antagonistic communities with dense positive connections in each community and sparse negative ones across communities. Unlike previous works \cite{bonchi2019discovering, tzeng2020discovering} that define polarization as a quality measure of graph partitions, our aim is to measure polarization based solely on graph structure. To achieve this, we apply a soft characterization of the polarized community structure in the graph based on signed random-walks. 

\begin{figure}[htbp]
    \centering
    \includegraphics[width=1\columnwidth]{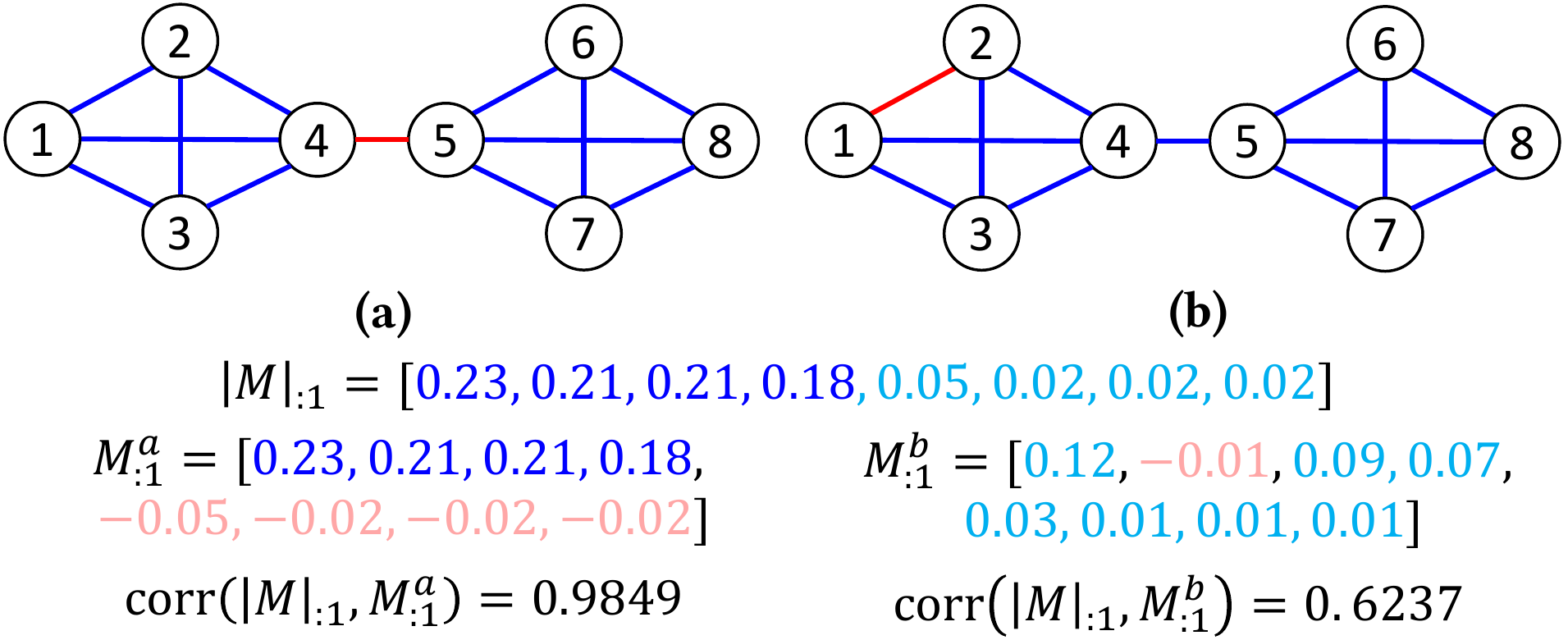}
    \caption{Illustration of our polarization measure based on signed random-walk dynamics ($t=3$). The transitions for unsigned ($|M|$) and signed ($M$) random-walks are more correlated in the polarized network (a), where a negative link connects two antagonistic communities. On the other hand, the correlation is lower in the less polarized network (b).}
    \label{fig::polarization-correlation}
\end{figure}

The key observation that supports our polarization measure is that polarization increases the correlation between unsigned and signed random-walk dynamics. 
Consider the two graphs shown in \autoref{fig::polarization-correlation}. While they share the same underlying topology, (a) is more polarized than (b) as its negative link connects the two communities. We then analyze how their signed random-walk transitions differ from unsigned ones. For node $1$ in graph (a), only signs of its inter-community transitions are changed, whose magnitudes are smaller, if not negligible, compared to the intra-community transitions. On the other hand, the negative link in the less polarized graph (b) significantly affects the signed intra-community transitions. As a consequence, the signed and unsigned transitions in the polarized graph are more correlated than in the less polarized one. 

We define the \emph{node-level polarization} as the (Pearson) correlation between a node's signed and unsigned random-walk transitions:
\begin{equation}
    \text{Pol}(u;t) = \corr(|M|_{:u}(t), M_{:u}(t))
\end{equation}
We can then define the \emph{graph-level polarization} as the mean node-level polarization for all nodes in the graph:
\begin{equation}
    \text{Pol}(G;t) = \mean_{u\in G}(\text{Pol}(u;t))
\end{equation}

The random-walk based polarization proposed here has two main advantages. First, it is partition-agnostic and does not depend on the availability of ground-truth communities or the outcome of community detection algorithms. Second, it can measure polarization at different structural scales by tuning the Markov time $t$---large $t$ captures polarization between macro-level communities (e.g., hostility between political parties) while small $t$ measures it at a micro scale (e.g., disagreement between factions within a party). 
In the next subsection, we will demonstrate that our measure is effective in characterizing both node and graph-level polarization.

\subsection{Polarization of Real-world Graphs}

\subsubsection{Node-level polarization}
\label{subsubsec::node_level_pol}
We apply our polarization measure to characterize nodes in a political network (\textsc{Congress}) \cite{thomas2006get}. Signed links in this network represent (un/)favorable interactions between U.S. congresspeople on the House floor in 2005. The statistics of the network are shown in \autoref{tab::datasets}. 

We rank all nodes by their polarization scores at a fixed $t=10$. The top 20 least and most polarized nodes are shown in \autoref{tab::least_polarized} and \autoref{tab::most_polarized} (both in Appendix \ref{subsec::app_pol}). The congressperson with the smallest score ($-0.6542$) is Henry Cuellar, a self-described moderate-centrist \cite{cuellar1}. As a Democrat, he voted with President Trump (Republican) nearly 75\% of the time, advocating his fellow Democrats to embrace a more conservative voting record \cite{cuellar2}. The second least polarized person is Jane Harman (with $-0.5376$). She was described as a centrist, particularly on defense and intelligence issues \cite{harman1}. She was also once called the best Republican in the Democratic Party \cite{harman2}. 
In general, we found that our polarization measure captures the political views (centrists vs. extremists) of the congresspeople based on their signed interactions.

\subsubsection{Graph-level polarization}
We also apply our graph-level polarization measure to characterize real-world signed graphs. In addition to \textsc{Congress}, we consider five other networks, with their statistics summarized in \autoref{tab::datasets}:
\begin{itemize}[leftmargin=*]
\item \textsc{WoW-EP8} \cite{kristof2020war}: Interaction network of editors in the eighth legislature of the European Parliament. Link signs indicate whether they collaborate or compete with each other. 
\item \textsc{Bitcoin-Alpha} and \textsc{Bitcoin-OTC} \cite{kumar2016edge}: Trust networks of Bitcoin traders on the platforms \emph{Bitcoin Alpha} and \emph{Bitcoin OTC}. Link weights and signs are based on users' rating of each other.
\item \textsc{Referendum} \cite{lai2018stance}: Interaction network of Twitter users on the 2016 constitutional referendum in Italy. Link signs are inferred from the stance of the users. 
\item \textsc{Wiki-RfA} \cite{west2014exploiting}: Voting network of Wikipedia users on adminship. Link signs indicate whether users vote for/against each other. 
\end{itemize}
\begin{table}[htbp]
  \centering
  \caption{An overview of the datasets.}
    \begin{tabular}{cccc}
    \toprule
          & $|\mathcal{V}|$     & $|\mathcal{E}|$  & $|\mathcal{E}_-|/|\mathcal{E}|$  \\
    \midrule
    \textsc{Congress} & 219  & 523 & 20.46\%\\
    \textsc{WoW-EP8}  & 789 & 116,009 & 18.63\%  \\
    \textsc{Bitcoin-Alpha} & 3,772 & 14,077 & 9.31\% \\
    \textsc{Bitcoin-OTC} & 5,872 & 21,431 & 14.71\% \\
    \textsc{Referendum} & 10,864 & 251,396 & 5.09\% \\
    \textsc{Wiki-RfA} & 11,275 & 169,925 & 22.04\%\\
    \bottomrule
    \end{tabular}%
  \label{tab::datasets}%
\end{table}%

As reference for graph-level polarization, we construct two synthetic graphs, one polarized and the other unpolarized, with the same underlying topology but different link signs, as shown in \autoref{fig::lfr}. The topology is based on the LFR benchmark \cite{lancichinetti2008benchmark} with the following parameters: two structural communities with 50 nodes, an average node degree of 12, and an inter-community link ratio of 0.15. Then, for the polarized version (\textsc{LFR-polarized}), we assign link signs fully based on the structural communities---negative for inter-community links and positive for intra-community links. For the unpolarized version (\textsc{LFR-unpolarized}), we first assign nodes to two random communities with the same cut size as structural communities. We then assign link signs based on the random communities following the same rule. In this way, both graphs have the same proportion of negative links and the same social balance---proportion of closed triangles in the graph that satisfy the social balance theory \cite{leskovec2010signed}---of 1.0. 

\begin{figure}[htbp]
    \centering
    \includegraphics[width=1\columnwidth]{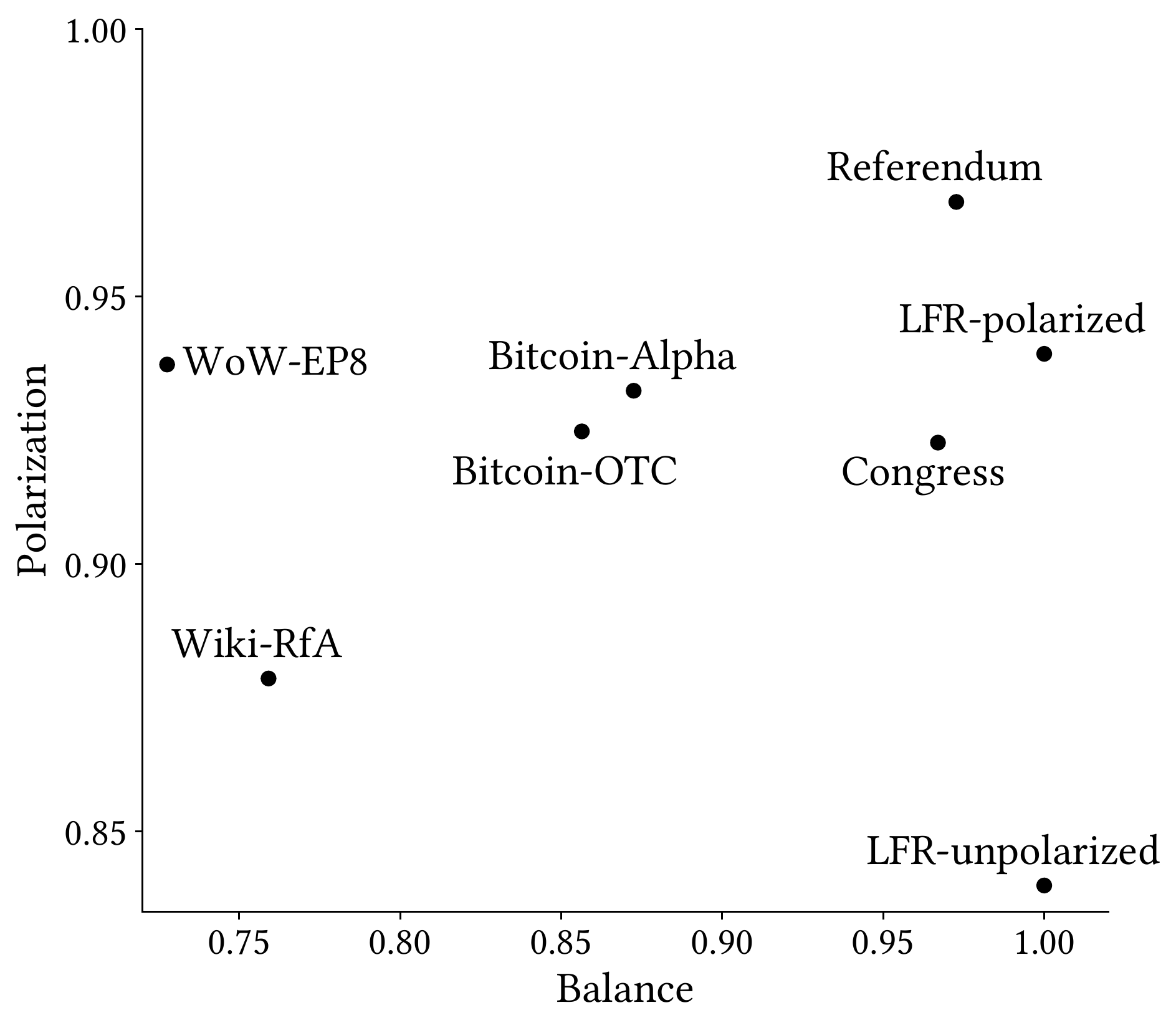}
    \caption{Polarization and social balance of real-world graphs, with reference to synthetic ones. Most real-world graphs are as polarized as the synthetic polarized one. }
    \label{fig::polarization-balance}
\end{figure}

The polarization and social balance for the real-world graphs along with the two synthetic ones are shown in \autoref{fig::polarization-balance}, with the Markov time $t$ selected based on signed link prediction performance (details in Section \ref{subsubsec::parameters}).
As we see, most real-world graphs are at the same level of polarization as \textsc{LFR-polarized}. On the other hand, their social balance is not directly related to polarization, contrary to as one might think \cite{garimella2017long}. 
The only dataset that falls behind in polarization is \textsc{Wiki-RfA}, which might be due to the more collaborative nature of its inter-community interactions. As we will discuss in the next section, effective link prediction for those polarized graphs requires a novel embedding scheme that satisfies \emph{polarized similarity consistency}.

\section{Polarized Embedding for Networks}
\label{sec::embedding}
Now that we have shown that many real-world graphs are polarized, here, we propose a novel embedding method for effectively predicting signed links in polarized graphs. We first demonstrate that existing embedding methods are incapable of this task due to their weak similarity consistency (see Section \ref{subsec::similarity_consistency}). 
We then introduce polarized embedding based on autocovariance and matrix factorization that addresses the limitations of existing approaches.
\begin{figure*}[htbp]
    \centering
    \includegraphics[width=\textwidth]{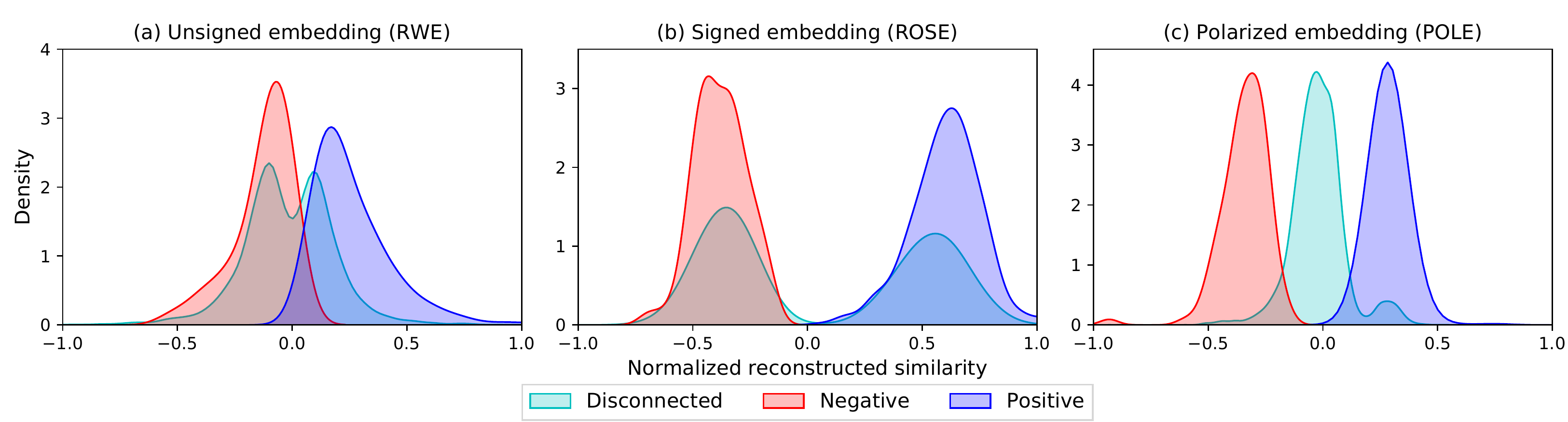}
    \caption{Distributions of the reconstructed similarity for different types of node pairs in \textsc{LFR-polarized}. Polarized embedding (c) enables separation of negatively connected pairs from the others while both (a) unsigned embedding (RWE \cite{random-walk-embedding}) and (b) signed embedding (ROSE \cite{javari2020rose}) fail to do so. }
    \label{fig::similarity_distribution}
\end{figure*}

\subsection{Limitation of Existing Methods}
The objective of signed link prediction \cite{beigi2019signed, xu2019link} is to predict both the existence and the signs of future links given the observed graph. Most of existing signed embedding methods focus on the second half of the task, sign prediction, which is specific to signed graphs. The first half, link prediction, is a common downstream task for unsigned embedding methods. Combining them seems like a viable solution for the task. However, this approach fails to predict negative links in polarized graphs because they are sparse and mostly appear as inter-community connections. They are hard, if not impossible, to be predicted by unsigned link prediction algorithms based on topology only. Without the existence of links known a priori, sign prediction itself is incapable of predicting negative links.

We identify the key weakness of existing embedding methods as that they only preserve weak similarity consistency. In particular, signed embedding methods \cite{wang2017signed, islam2018signet, kim2018side, chen2018bridge, xu2019link, javari2020rose} just ensure \emph{signed similarity consistency}---positively related pairs are separable from negatively related pairs---for sign prediction. And similarly, unsigned embedding methods \cite{perozzi2014deepwalk, grover2016node2vec, random-walk-embedding} only guarantee the \emph{topological similarity consistency}---topologically close node pairs are separable from distant pairs---for link prediction. While positive links are detectable as the intersection of positive and topologically close pairs, negative links in polarized graphs would remain hidden with other topologically distant pairs. \autoref{fig::similarity_distribution} (a) and \autoref{fig::similarity_distribution} (b) illustrate this idea. They show the distributions of reconstructed similarity of positively connected pairs, negatively connected pairs, and disconnected pairs for \textsc{LFR-polarized} via dot products of unsigned and signed embedding, respectively. While both embedding methods are able to capture their respective similarity consistency, negative links are always inseparable from disconnected pairs. This motivates us to design an embedding method with a stronger similarity consistency that enables the separation of negative pairs from the others. We describe such an embedding method next.

\subsection{The Solution: Polarized Embedding}
Our solution to the separability of negative pairs is polarized embedding, a novel embedding scheme that captures \emph{polarized similarity consistency}, the strongest consistency among the three. This consistency guarantees that negatively related pairs are more dissimilar than unrelated topologically distant pairs (thus ``polarized''), which are more dissimilar than positively related pairs. In this way, the negatively related pairs stand out at the other end of the similarity spectrum, easily separable from others. In the following subsections, we introduce our polarized embedding method, POLE, that applies signed autocovariance similarity and matrix factorization.

\subsubsection{Signed autocovariance similarity}
POLE is based on the signed random-walk introduced in Section \ref{subsec::signed_random_walk}. As discussed in Section \ref{subsec::similarity_consistency}, both the entries and the dot products of columns of the signed random-walk transition matrix $M(t)$ can be viewed as similarity metrics that satisfy \emph{polarized similarity consistency}. Therefore, one can directly use a low-rank representation of them---for example, the matrix factorization of $M(t)$ or $M(t)^TM(t)$---as embeddings. However, those embeddings may not be effective for link-level inference. Instead, we propose to take advantage of the random-walk based similarity metrics that have been well understood in unsigned embedding. In particular, it has been shown that autocovariance similarity \cite{delvenne2010stability} enables state-of-the-art performance in unsigned link prediction by incorporating node degree information \cite{random-walk-embedding}. This coincides with our goal of predicting signed links in polarized graphs. Thus, we first extend autocovariance similarity to signed graphs.

The unsigned autocovariance similarity is built upon the co-visiting probability for node pairs in a walk. However, as defined in our signed random-walk, $M(t)$ is not a stochastic matrix and does not encode probabilities. Instead, we resort to another interpretation of autocovariance---a centered dynamic similarity metric \cite{schaub2019multiscale}. The dynamic similarity based on $M(t)$ is
\begin{equation}
    R(t) = M(t)^T W M(t)
    \label{eqn::autocovariance}
\end{equation}
where $W \in \mathbb{R}^{n\times n}$ is a weight matrix. In the unsigned case, selecting $W=\Pi - \pi \pi^T$ makes $R(t)$ equivalent to the autocovariance similarity, where $\pi \in \mathbb{R}^n$ is the stationary distribution of the unsigned random-walk and $\Pi = \diag(\pi)$. 
For the signed case, $M(t)$ is singular if and only if the graph is perfectly balanced \cite{kunegis2010spectral}. This means that there is no ``stationary distribution'' for signed random-walks on most real-world graphs. However, as \cite{lambiotte2014random} points out, the role of stationary distribution in the similarity formulation is to provide a centrality measure---in the unsigned case, the degree centrality. Thus, we can also use node degrees to construct the weight matrix:
\begin{equation}
    W = \frac{1}{\vol(G)}D - \frac{1}{\vol(G)^2}dd^T
\end{equation}
where $\vol(G) = \sum_u d_u$. And substituting this weight matrix into \autoref{eqn::autocovariance} leads to the signed autocovariance similarity matrix. To the best our knowledge, we are the first to extend random-walk based similarity metrics to signed graphs in a principled manner. 

\subsubsection{Matrix factorization}
We are now ready to introduce how to generate embedding based on signed autocovariance. Let $\mathbf{u}_u \in \mathbb{R}^k$ be the embedding of node $u$ and $U = (\mathbf{u}_1, \dotsc, \mathbf{u}_n)^T \in \mathbb{R}^{n\times k}$ be the embedding matrix. We use the dot product in the embedding space to preserve the signed autocovariance similarity $R$:
\begin{equation}
\begin{aligned}
U^* &= \argmin_U \sum_{u,v} (\mathbf{u}_u^T\mathbf{u}_v^{} - R_{uv})^2\\
&= \argmin_U \norm{UU^T - R}_F^2
\end{aligned}
\label{eqn::obj_und_embed}
\end{equation}
This leads to a straightforward matrix factorization algorithm to find the optimal embedding. Specifically, $U^*=Q_k\sqrt{\Lambda_k}$---where $R=Q\Lambda Q^T$ is the Singular Value Decomposition (SVD) of $R$---is the optimal solution of $U$ under the constraint $\rank(UU^T) = k$ \cite{eckart1936approximation}. 

\autoref{fig::similarity_distribution} (c) shows distributions of reconstructed similarity of different types of node pairs for polarized embedding. It is clear that negative pairs can be effectively separated from unrelated and positive pairs. In addition, with negative pairs staying at the negative end of the similarity spectrum, positive pairs are also more separable. This demonstrates the strength of our polarized embedding that captures \emph{polarized similarity consistency}.

\section{Experiments}
\label{sec::experiments}
In this section, we demonstrate the effectiveness of POLE in signed link prediction using the six datasets in \autoref{tab::datasets}. Our code is available at \url{https://github.com/zexihuang/POLE}.
\subsection{Experimental Settings}
\subsubsection{Baselines}
We consider the following baselines:
\begin{itemize}[leftmargin=*]
    \item SiNE \cite{wang2017signed}: Combines an objective function based on social balance theory with a learned pairwise similarity function. 
    \item SIGNet \cite{islam2018signet}: Attempts to capture higher-order balance structure by computing embeddings for which dot product approximates the signed proximity for links in the graph. 
    \item SIDE \cite{kim2018side}: Extends random-walk based embedding to signed graphs. Embeddings are learned based on maximum likelihood using pairwise proximities that are sensitive to link signs.
    \item BESIDE \cite{chen2018bridge}: Applies balance and status theory to model signed triangles and ``bridge'' links, which are not included in a triangle. 
    \item SLF \cite{xu2019link}: Decomposes node embeddings into two types of latent factors (positive and negative). These factors, which are learned via coordinate descent, are applied to generate four types of scores corresponding to positive, negative, neutral, and no link. 
    \item ROSE \cite{javari2020rose}: Transforms the signed network into an unsigned bipartite one by representing each node multiple times. Embeddings for the unsigned network are generated using a  random-walk based approach and combined into signed embeddings. 
\end{itemize}

\subsubsection{Downstream task}
\label{subsubsec::downstream_task}
We focus on signed link prediction, which consists of predicting of both link existence and signs. We randomly remove 20\% of links while ensuring that the residual graph is connected and embed the residual graph. Then, for POLE, we rank all disconnected node pairs based on reconstructed dot product similarity and predict the most similar/dissimilar pairs as positive/negative links. For baselines, we follow \cite{xu2019link} and train two logistic regression classifiers on concatenated node embeddings for positive/negative pairs vs disconnected pairs, respectively. While ranking by classifier scores is adopted by most baselines, we also conduct experiments with baselines using the same dot product similarity ranking as for POLE. The results are similar and included in Appendix \ref{subsec::app_slp}. 
We report \emph{precision@k} \cite{lu2011link} for positive/negative links respectively, where $k$ is the number of top pairs in terms of the ratio of removed positive/negative links, ranging from 10\% to 100\%.

Since the baselines are designed to only capture \emph{signed similarity consistency} and thus may perform poorly without link existence information, we also consider supplementing them with an unsigned embedding method. This also allows us to analyze the interaction between unsigned similarity and the signed similarity captured by our polarized embedding. Specifically, we apply RWE \cite{random-walk-embedding} with unsigned autocovariance and continuous random-walk and compute the reconstructed similarity based on embeddings for the unsigned residual graph. 
We then train logistic regression classifiers to combine the unsigned similarity (encoding link existence information) and either reconstructed similarity (for POLE) or classifier scores (for baselines) for final ranking and report \emph{precision@k} scores. 

\subsubsection{Parameters}
\label{subsubsec::parameters}
We set the number of embedding dimensions $k$ to 40 for all methods (except ROSE which requires a multiple of 3, we set it to 42). The only other parameter in RWE and POLE is the Markov time $t$, which is selected from \{$10^{0.0}, 10^{0.1}, \dotsc, 10^{1.0}$\} based on signed/unsigned link prediction. Other parameters for baselines are set as recommended in the original papers.

\subsection{Results}
\begin{figure*}[htbp]
    \centering
    \includegraphics[width=\textwidth]{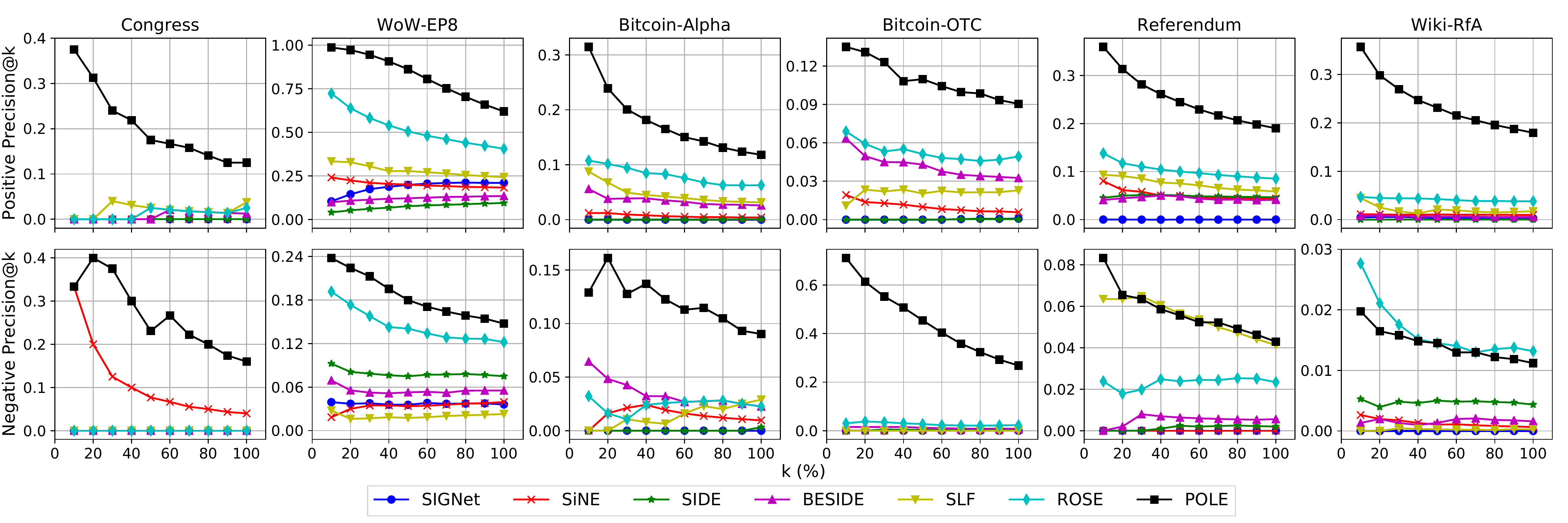}
    \caption{Comparison of performance of signed link prediction between POLE and baselines. POLE outperforms all baselines in all datasets on both positive and negative link prediction, except for negative links in \textsc{Wiki-RfA}, the least polarized network. }
    \label{fig::signed_link_prediction}
\end{figure*}
\begin{figure*}[htbp]
    \centering
    \includegraphics[width=\textwidth]{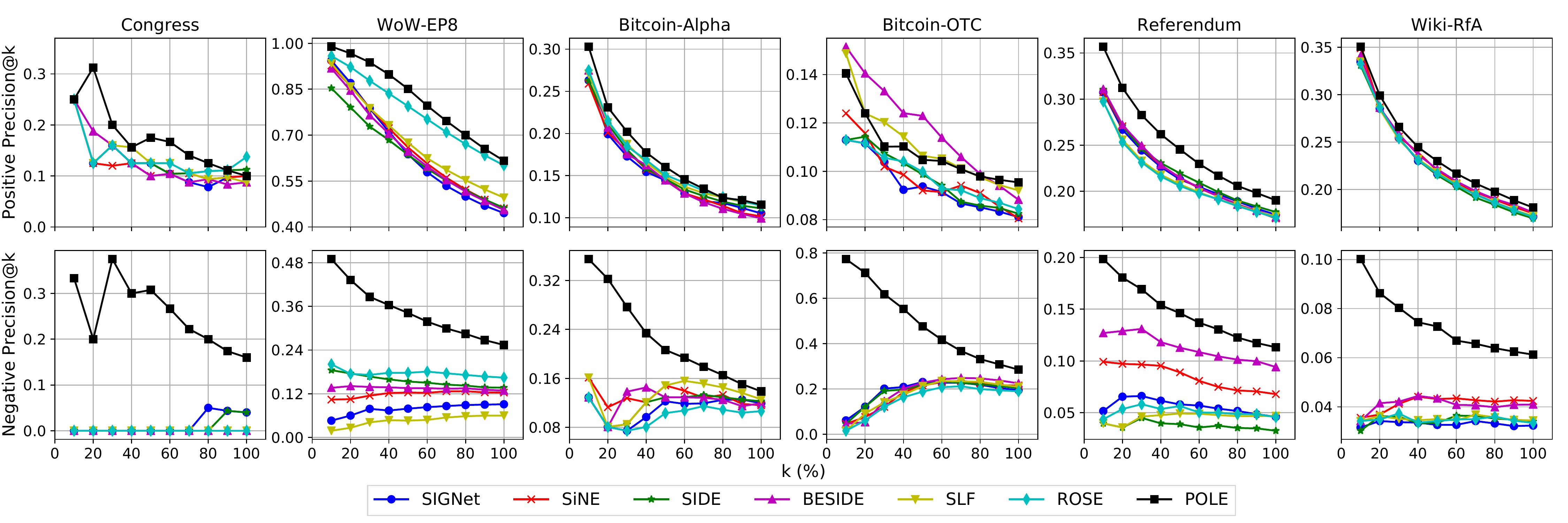}
    \caption{Comparison of performance of signed link prediction with link existence information between POLE and baselines. While adding unsigned similarity narrows the performance gap between POLE and baselines on positive link prediction, it significantly improves POLE on negative link prediction and keeps its edge. }
    \label{fig::signed_link_prediction_with_unsigned}
\end{figure*}
\subsubsection{Signed link prediction}
\autoref{fig::signed_link_prediction} shows the signed link prediction performance for different methods without supplementing the unsigned link existence information. We first note that the proposed method, POLE, outperforms all baselines in almost all datasets for both positive and negative links. The average gains in \emph{precision@k} over the best baseline for each dataset are 629.2\%/58.8\%/115.7\% /108.3\% /142.5\%/468.8\% on positive links and 220.0\%/27.6\%/261.8\% /1539.4\%/4.3\%/-10.9\% on negative links. 
This shows that POLE enables state-of-the-art signed link prediction performance. 

While the main motivation for POLE is predicting negative links, its superior performance in positive link prediction is also consistent with our previous similarity analysis (see \autoref{fig::similarity_distribution} (b) and \autoref{fig::similarity_distribution} (c)). 
The only scenario where POLE underperforms the best baseline is the negative link prediction on \textsc{Wiki-RfA}. This is actually not surprising as \textsc{Wiki-RfA} is the least polarized network among the six, as shown in \autoref{fig::polarization-balance}. When the graph is less polarized (e.g., \autoref{fig::lfr} (b)), negative links are formed independently from community structure, and therefore the \emph{polarized similarity consistency} may not be the best criterion to embed nodes.

\subsubsection{Signed link prediction with link existence information}
\label{subsubsec::slp_link_existence}
We now supplement each method with unsigned link existence information. Results are shown in \autoref{fig::signed_link_prediction_with_unsigned}. Adding unsigned information narrows down the average performance gains of POLE over baselines on positive links to 18.0\%/5.1\%/4.4\%/-6.6\%/11.2\%/3.5\%. But our method still exhibits significant gains on negative links, outperforming the best baseline by 300.0\%/93.9\%/70.4\%/243.8\%/29.6\%/79.7\% for all datasets including the least polarized \textsc{Wiki-RfA}. Even with unsigned link existence information, our polarized embedding method has an edge over existing approaches, especially for negative links.

\begin{figure*}[htbp]
    \centering
    \includegraphics[width=\textwidth]{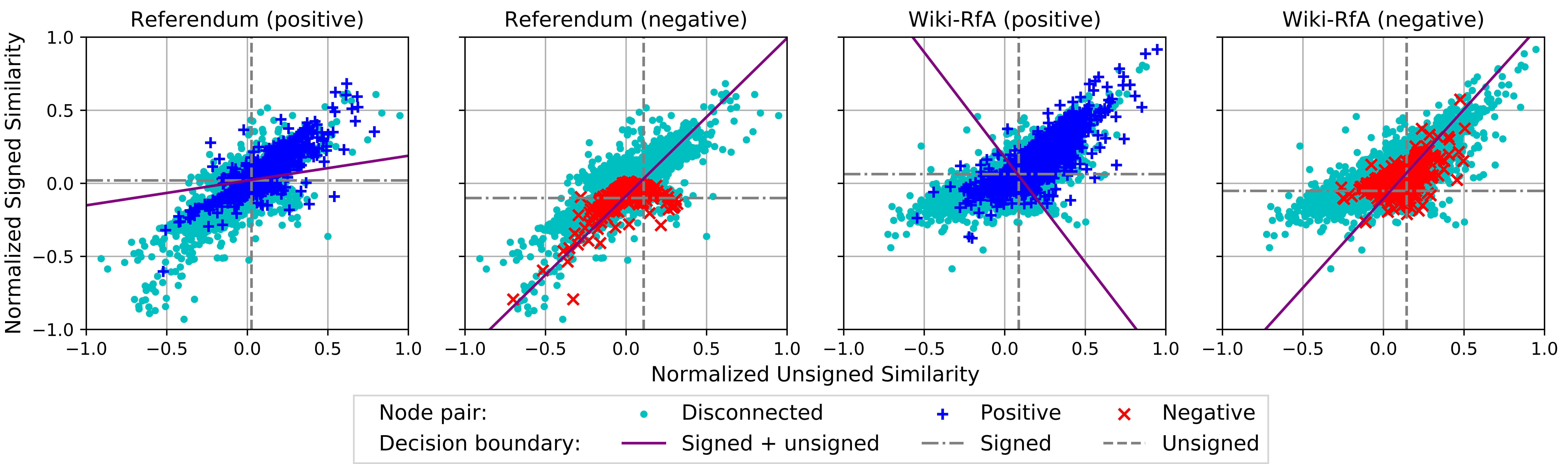}
    \caption{Scatter plot of the reconstructed signed and unsigned similarity for different node pairs in signed link prediction, along with the decision boundaries based on each similarity and a combination of both (via the classifier). Combining signed and unsigned similarity improves prediction for negative links but has a negligible effect on predicting positive links. }
    \label{fig::classifier_visualization}
\end{figure*}

The notable improvement of baselines after adding unsigned link existence information is consistent with our expectation. For positive links, combining the signed/unsigned embeddings enables the separation of positive pairs from negative/disconnected pairs, respectively, as illustrated in \autoref{fig::similarity_distribution} (a) and \autoref{fig::similarity_distribution} (b). This boosts baselines to a performance comparable with ours. For negative links, while all baselines still underperform POLE by a large margin, their predictions are improved significantly compared to the scenario without link existence information. The most prominent improvement is observed for \textsc{Bitcoin-OTC}, from negligible precision scores to notable ones of around 0.2 at $k=100\%$. This is reasonable since none of the networks are fully polarized, and not all negative links in the networks are inter-community. Those that are intra-community can be correctly predicted by unsigned similarity first and then correct signs can be discovered using the baselines. The inter-community negative links, however, remain exclusively predictable via \emph{polarized similarity consistency}, for which our polarized embedding method is designed.

At this point, it is important to analyze the interaction between unsigned similarity and our polarized embedding. While the performance for negative links more than doubles on average across the datasets, precision for positive links decreases by an average of 2.9\%. To understand this phenomenon, we draw scatter plots of the reconstructed signed and unsigned similarity for different node pairs in signed link prediction, along with the decision boundaries based on each similarity and a combination of both (via the classifier). The results for \textsc{Referendum} and \textsc{Wiki-RfA} are shown in \autoref{fig::classifier_visualization}, with the rest in \autoref{fig::classifier_visualization_appendix} in Appendix \ref{subsec::app_cls}. For positive pairs, signed and unsigned similarities are highly correlated, and signed similarity alone is predictive enough. This is consistent with the fact that regions above the signed and combined decision boundaries highly overlap. On the other hand, prediction for negative links is clearly improved by combining signed and unsigned similarities. The decision boundaries of the learned classifiers imply that negative pairs should have smaller signed similarity than unsigned similarity. This rule is especially useful as it filters out a large number of disconnected pairs with large negative signed similarity but even larger negative unsigned similarity. 
Finally, this plot is another illustration of how our method captures polarization. 
\textsc{Referendum}, the most polarized graph (see \autoref{fig::polarization-balance}) of all, shows a clearer separation between positive and negative pairs compared to \textsc{Wiki-RfA}, which is the least polarized.

\section{Related Work}
\label{sec::related_work}
\noindent\textbf{Measuring polarization.} Social polarization has received great attention with the increasing popularity in online social media \cite{guerra2013measure}. In the context of networks/graphs, polarization is often defined as the cohesiveness of communities. Several community quality measures,
such as conductance \cite{kannan2004clusterings} and modularity \cite{newman2006modularity}, have been used for analyzing polarization in the U.S. congress \cite{waugh2011party} and online social media platforms \cite{conover2011political}. Later works consider other measures based on the community structure. For example, \cite{guerra2013measure} measures polarization based on the concentration of high-degree nodes in the community boundaries, and \cite{garimella2018quantifying} extends it with betweenness and inter-community random-walk transitions. Likewise, polarization in signed graphs is often related to the objective function of signed community detection. In \cite{bonchi2019discovering}, the authors define polarization as the size-normalized signed cut between two conflicting communities, which is later extended to multiple communities in \cite{tzeng2020discovering}. The main distinction of our polarization measure compared with existing methods is that it leverages the signed random-walk to capture the community structure at multiple scales and does not require specific community assignments as input. 

\noindent\textbf{Signed link prediction.} Link prediction is a common inference task for graphs \cite{lu2011link, martinez2016survey}. Link signs in signed graphs add another dimension to the problem, leading to the different tasks of (1) link sign prediction \cite{leskovec2010predicting, javari2014cluster, song2015link}, (2) link existence prediction \cite{yuan2019graph, javari2020rose}, and (3) signed link prediction \cite{chiang2011exploiting, hsieh2012low, wang2017online}. Signed link prediction is the most challenging one among them, requiring both existence and signs of links to be inferred. Existing signed link prediction methods are based on feature engineering \cite{chiang2011exploiting}, matrix factorization \cite{hsieh2012low, ye2013predicting}, and graph embedding \cite{xu2019link}. Due to sparsity of links, several methods also consider side information, including user-item ratings \cite{beigi2019signed}, user-user opinions \cite{beigi2016exploiting, beigi2019signed}, and user-review ratings \cite{tang2014predictability}. Our work focuses on signed link prediction in polarized graphs, which is an even harder problem as negative links mostly appear as sparse inter-community connections. Yet by incorporating polarization in the embedding, our method enables state-of-the-art signed link prediction without using additional information.

\noindent\textbf{Signed graph embedding.} 
Signed graph embedding enables the application of classical machine learning algorithms to graph-based downstream tasks, such as signed link prediction \cite{chiang2011exploiting, hsieh2012low}, node classification \cite{islam2018signet}, and polarized community detection \cite{bonchi2019discovering, tzeng2020discovering}. Existing methods are often based on extending unsigned embedding methods (such as those based on random-walks \cite{perozzi2014deepwalk, grover2016node2vec}) and incorporating social theories (such as social balance theory \cite{heider1946attitudes} and social status theory \cite{guha2004propagation}). Both SNE \cite{yuan2017sne} and SIDE \cite{kim2018side} extend random-walk based objective functions to signed embedding. SiNE \cite{wang2017signed} incorporates social balance theory and uses a learned pairwise similarity function modeled as a multi-layer neural network. SIGNet \cite{islam2018signet} captures higher-order balance structure with efficient sampling algorithms. BESIDE \cite{chen2018bridge} combines social balance theory and social status theory to model triangles and ``bridge'' links for embedding. SLF \cite{xu2019link} models each node with four signed latent factors in order to capture positive, negative, neutral, and the absence of a relationship between them. ROSE \cite{javari2020rose} transforms the signed network into a bipartite unsigned network and leverages an existing random-walk based method \cite{grover2016node2vec} for embedding. While these methods capture the \emph{signed similarity consistency} needed for sign prediction (see Figure \ref{fig::similarity_distribution}), they are unable to predict negative links in polarized graphs even when combined with unsigned embedding \cite{random-walk-embedding}. By comparison, our polarized embedding preserves \emph{polarized similarity consistency} and can effectively predict both positive and negative links in real-world signed graphs.

\section{Conclusion}
\label{sec::conclusions}
We have introduced several analytical tools for understanding and combating polarization in signed graphs. They include a partition-agnostic polarization measure for both nodes and graphs and an embedding algorithm that enables state-of-the-art signed link prediction, especially for the hostile links in polarized graphs. We hope that our work will be an important step towards making social media an environment for healthy and constructive communication among individuals with diverse viewpoints.

\begin{acks}
This work is partially funded by NSF via grant IIS 1817046 and DTRA via grant HDTRA1-19-1-0017.
\end{acks}

\clearpage
\bibliographystyle{ACM-Reference-Format}
\bibliography{reference}
\clearpage
\appendix
\onecolumn
\section{Appendix}
\subsection{Polarization Scores for the U.S. Congresspeople}
\label{subsec::app_pol}
\autoref{tab::least_polarized} and \autoref{tab::most_polarized} (both referred in Section \ref{subsubsec::node_level_pol}) show the polarization scores for the top 20 least and most U.S. congresspeople. 
\begin{table}[htbp]
\begin{multicols}{2}
  \centering
  \caption{Top 20 least polarized members of the U.S. Congress in terms of our random-walk based polarization measure. }
    \begin{tabular}{cccc}
    \toprule
    Congressperson & State & Party & Score \\
    \midrule
    Henry Cuellar & Texas & D     & -0.6542 \\
    Jane Harman & California & D     & -0.5376 \\
    Curt Weldon & Pennsylvania & R     & -0.4381 \\
    Dutch Ruppersberger & Maryland & D     & -0.4318 \\
    Jim Moran & Virginia & D     & -0.3832 \\
    Dave Obey & Wisconsin & D     & -0.3588 \\
    Wayne Gilchrest & Maryland & R     & -0.3503 \\
    Duke Cunningham & California & R     & -0.3248 \\
    Al Edwards & Texas & D     & -0.3063 \\
    Lincoln Davis & Tennessee & D     & -0.2901 \\
    Joe Barton & Texas & R     & -0.1492 \\
    Charles Bass & New Hampshire & R     & -0.1381 \\
    Mary Bono & California & R     & -0.1381 \\
    Charlie Dent & Pennsylvania & R     & -0.1381 \\
    Nick Rahall & West Virginia & D     & -0.1357 \\
    Bill Young & Florida & R     & -0.1250 \\
    Roger Wicker & Mississippi & R     & -0.1189 \\
    Charles Rangel & New York & D     & -0.0926 \\
    Candice Miller & Michigan & R     & -0.0743 \\
    Jim Kolbe & Arizona & R     & -0.0716 \\
    \bottomrule
    \end{tabular}%
 \label{tab::least_polarized}%
  \centering
  \caption{Top 20 most polarized members of the U.S. Congress in terms of our random-walk based polarization measure. }
    \begin{tabular}{cccc}
    \toprule
    Congressperson & State & Party & Score \\
    \midrule
    Steve Buyer & Indiana & R     & 0.9897 \\
    James T. Walsh & New York & R     & 0.9877 \\
    Charles H. Taylor & North Carolina & R     & 0.9851 \\
    Zach Wamp & Tennessee & R     & 0.9776 \\
    Tom Cole & Oklahoma & R     & 0.9747 \\
    Geoff Davis & Kentucky & R     & 0.9261 \\
    Steve Israel & New York & D     & 0.9019 \\
    Marsha Blackburn & Tennessee & R     & 0.8875 \\
    Virginia Foxx & North Carolina & R     & 0.8814 \\
    Steve King & Iowa  & R     & 0.8516 \\
    Joe Crowley & New York & D     & 0.8444 \\
    Virgil Goode & Virginia & R     & 0.8175 \\
    Robert E. Cramer & Alabama & D     & 0.8117 \\
    Rubén Hinojosa & Texas & D     & 0.8030 \\
    Carolyn McCarthy & New York & D     & 0.8030 \\
    David Dreier & California & R     & 0.7874 \\
    Christopher Cox & California & R     & 0.7868 \\
    Jo Ann Davis & Virginia & R     & 0.7867 \\
    John E. Peterson & Pennsylvania & R     & 0.7867 \\
    Ray LaHood & Illinois & R     & 0.7793 \\
    \bottomrule
    \end{tabular}%
  \label{tab::most_polarized}
 \end{multicols}
\end{table}%

\subsection{Signed Link Prediction for Baselines with Reconstructed Similarity Ranking}
\label{subsec::app_slp}
\begin{figure}[htbp]
    \centering
    \includegraphics[width=\textwidth]{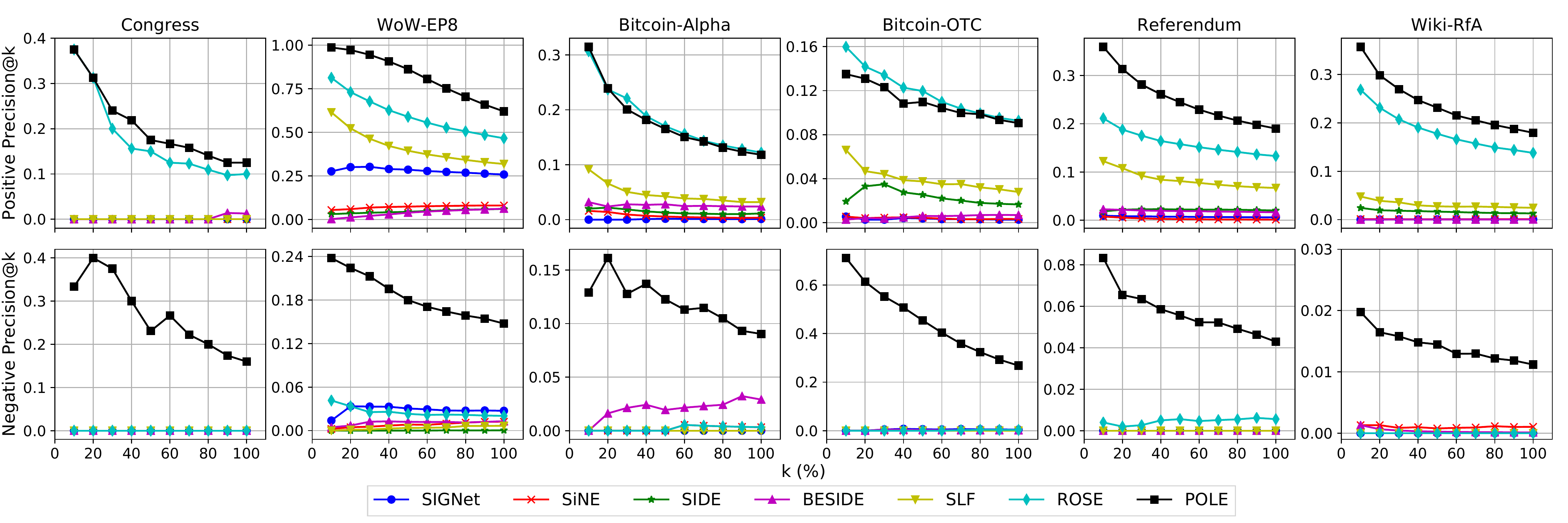}
    \caption{Comparison of performance of signed link prediction between POLE and baselines with reconstructed similarity ranking. POLE outperforms all baselines in almost all datasets on both positive and negative link prediction.}
    \label{fig::signed_link_prediction_linear}
\end{figure}

\begin{figure}[htbp]
    \centering
    \includegraphics[width=\textwidth]{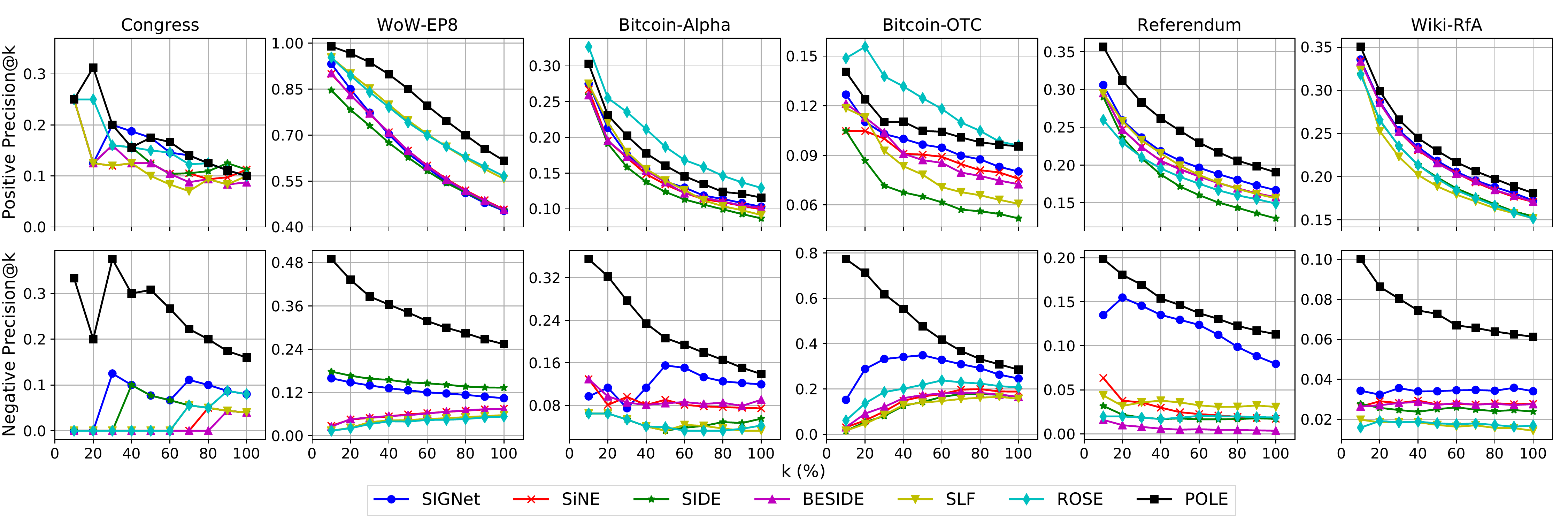}
    \caption{Comparison of performance of signed link prediction with link existence information between POLE and baselines with reconstructed similarity ranking. While adding unsigned similarity narrows the performance gap between POLE and baselines on positive link prediction, it significantly improves POLE on negative link prediction and keeps its edge. }
    \label{fig::signed_link_prediction_with_unsigned_linear}
\end{figure}

\autoref{fig::signed_link_prediction_linear} and \autoref{fig::signed_link_prediction_with_unsigned_linear} (both referred in Section \ref{subsubsec::downstream_task}) show the signed link prediction performance comparison between POLE and baselines with reconstructed similarity ranking via embedding dot products. 

\subsection{Scatter Plots of Signed and Unsigned Similarities with Decision Boundaries}
\label{subsec::app_cls}
\begin{figure}[htbp]
    \centering
    \includegraphics[width=\textwidth]{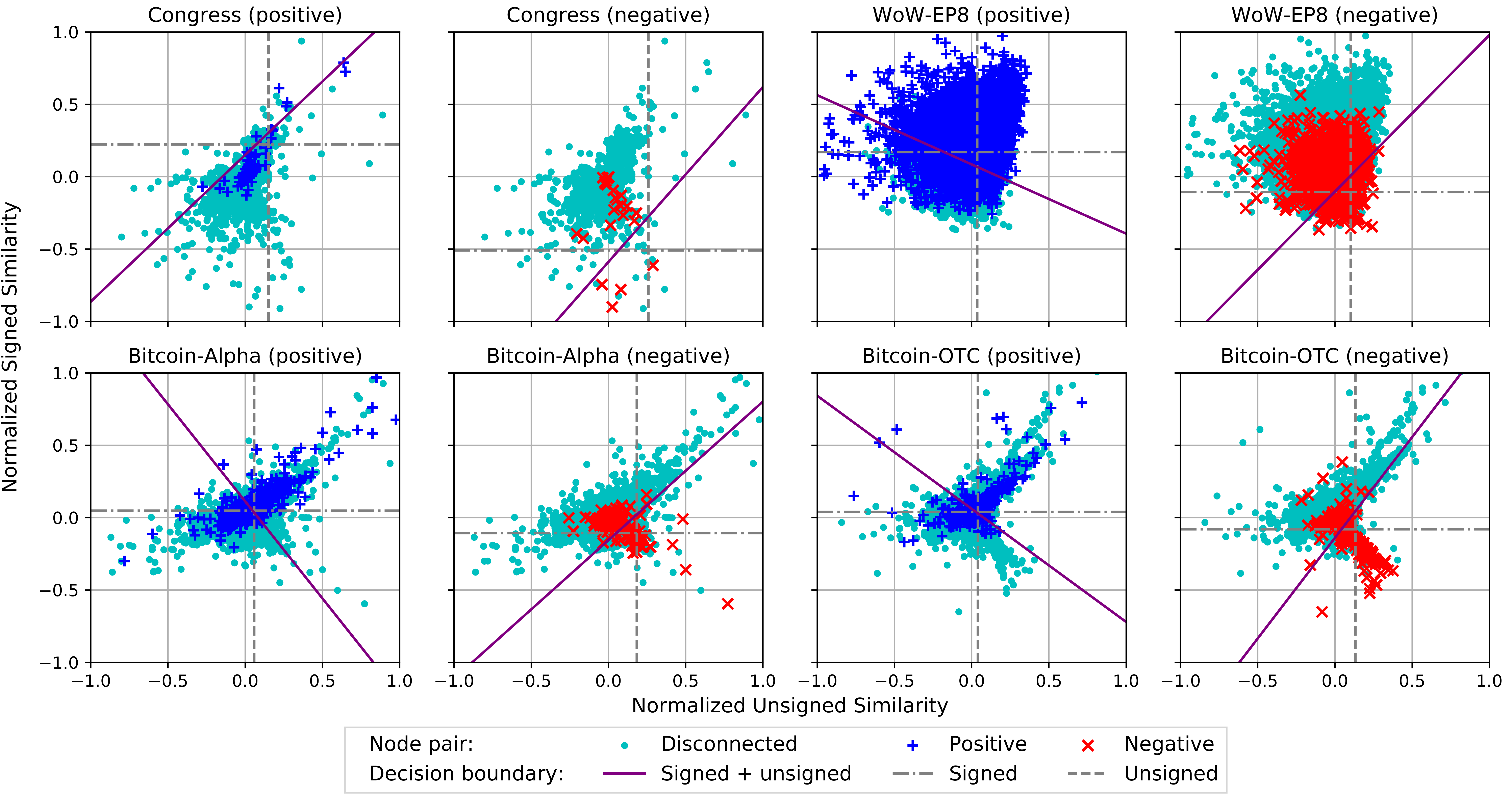}
    \caption{Scatter plot of the reconstructed signed and unsigned similarity for different node pairs in signed link prediction, along with the decision boundaries based on each similarity and a combination of both (via the classifier). Combining signed and unsigned similarity improves prediction for negative links but has a negligible effect on predicting positive links. }
    \label{fig::classifier_visualization_appendix}
\end{figure}

\autoref{fig::classifier_visualization_appendix} (referred in Section \ref{subsubsec::slp_link_existence}) shows the scatter plots of signed and unsigned similarity for different node pairs in signed link prediction along with the decision boundaries for \textsc{Congress}, \textsc{WoW-EP8}, \textsc{Bitcoin-Alpha}, and \textsc{Bitcoin-OTC}, respectively. 

\end{document}